\documentclass[
reprint,
bibnotes,
amsmath,amssymb,
aps,
prd
]{revtex4-2}

\usepackage{graphicx}
\usepackage{dcolumn}
\usepackage{enumitem}
\usepackage{bm}
\usepackage{float}
\usepackage[many]{tcolorbox}
\usepackage{shuffle}
\usepackage{environ}
\usepackage{physics}
\usepackage{tabularx}
\usepackage{diagbox}
\usepackage{subcaption}
\usepackage{multirow}

\newcommand*\diff{\mathop{}\!\mathrm{d}}
\usepackage{CJK}

\begin{document}

\begin{CJK*}{UTF8}{}
\CJKfamily{gbsn}

\title{Surface Kinematics and ``The'' Yang-Mills Integrand}

\author{Nima Arkani-Hamed$^{1}$}
\email{arkani@ias.edu}
\author{Qu Cao(曹 趣)$^{2,3}$}
\email{qucao@zju.edu.cn}
\author{Jin Dong(董 晋)$^{2,4}$}
\email{dongjin@itp.ac.cn}
\author{Carolina Figueiredo$^{5}$}
\email{cfigueiredo@princeton.edu}
\author{Song He(何 颂)$^{2,6,7}$}
\email{songhe@itp.ac.cn}

\affiliation{$^{1}$School of Natural Sciences, Institute for Advanced Study, Princeton, NJ, 08540, USA \\
$^{2}$Institute of Theoretical Physics, Chinese Academy of Sciences, Beijing 100190, China \\
$^{3}$Zhejiang Institute of Modern Physics, Department of Physics, Zhejiang University, Hangzhou, 310027, China \\
$^{4}$School of Physical Sciences, University of Chinese Academy of Sciences, No.19A Yuquan Road, Beijing 100049, China \\
$^{5}$Jadwin Hall, Princeton University, Princeton, NJ, 08540, USA \\
$^{6}$School of Fundamental Physics and Mathematical Sciences, Hangzhou Institute for Advanced Study and ICTP-AP, UCAS, Hangzhou 310024, China\\
$^{7}$Peng Huanwu Center for Fundamental Theory, Hefei, Anhui 230026, P. R. China}

\begin{abstract}
It has been a long-standing challenge to define a canonical loop integrand for non-supersymmetric gluon scattering amplitudes in the planar limit. Naive integrands are inflicted with $1/0$ ambiguities associated with tadpoles and massless external bubbles, which destroy integrand-level gauge invariance as well as consistent on-shell factorization on single loop-cuts. In this letter, we show that this essentially kinematical obstruction to defining ``the'' integrand for Yang-Mills theory has a structural solution, handed to us by the formulation of gluon amplitudes in terms of curves on surfaces. This defines ``surface kinematics'' generalizing momenta, making it possible to define ``the'' integrand satisfying both a (surface generalized) notion of gauge-invariance and consistent loop-cuts. The integrand also vanishes at infinity in appropriate directions, allowing it to be recursively computed for non-supersymmetric Yang-Mills theory in any number of dimensions. We illustrate these ideas through one loop for all multiplicity, and for the simplest two-loop integrand. 
 
\end{abstract}
\maketitle
\end{CJK*}

\noindent {\bf Introduction.}---
Scattering amplitudes are most powerfully understood, constrained and sometimes entirely determined by a systematic understanding of their singularities. 
At tree-level, amplitudes are rational functions with simple poles; the residue on these poles factorizes into the product of lower-point amplitudes. This is both a defining feature of tree amplitudes and has important practical consequences: taken together with non-trivial control of the behavior of the amplitude when approaching infinity in various directions ({\it c.f.}~\cite{Arkani-Hamed:2008bsc}), factorization allows for a recursive computation of tree-amplitudes, most famously through the BCFW recursion relations for gauge theories and gravity \cite{BCF,BCFW}. 

At loop-level, integrated amplitudes have much more interesting singularities associated with transcendental functions, but at least in the planar limit for theories with color, there is a good notion of a planar integrand \cite{AllLoopN4}, which is again simply a rational function of external and loop momenta. It is thus natural to fully characterize the pole structure and factorization patterns for loop-integrands, beginning with understanding ``single-loop-cuts'' where a single loop is put on-shell. 

However, single-cuts have infamous pathologies and are not well-defined. At one-loop, a single-cut where a loop propagator $\ell^2 \to 0$ should be given by the ``forward limit'' of a tree-amplitude~\cite{Feynman:1963ax}, where two color-adjacent particles are added with equal and opposite momenta $p_1^\mu =  \ell^\mu, p_2^\mu = -\ell^\mu$. This is a degenerate kinematic configuration, where the tree has a  $1/(p_1 + p_2)^2 = 1/0$ singularity from cutting open one-loop tadpoles. The same issue arises with external bubbles, which also produce a $1/0$ factor for on-shell external states.  In practice, tadpoles and external bubbles can be manually removed from the one-loop integrand, but this destroys the factorization properties, no longer matching tree amplitudes in the forward limit.  
These artificial integrands are also not on-shell gauge-invariant under 
shifting the external polarizations $\epsilon_j^\mu \to \epsilon_j^\mu + \alpha p_j^\mu$~\footnote{Of course, since all the difficulties are associated with tadpoles and massless external bubbles, the gauge-non-invariance is also proportional to tadpoles and external bubbles, which are scaleless-integrands that integrate to zero, so there is no difficulty with gauge-invariance for the integrated amplitude. But the integrand itself is not a good object.}.

It is a happy feature of planar ${\cal N}=4$ SYM that the super-sum over states removes the $1/0$ singularities so the forward limit is well-defined~\cite{Simon_TreeLoop}. This gives a canonical notion of ``the'' integrand, that can be determined using all-loop recursion~\cite{AllLoopN4}, and also the object computed by combinatorial-positive-geometric ideas of the amplituhedron~\cite{Amplituhedron}. But the absence of a canonical notion of ``the'' integrand for non-supersymmetric theories of the real world has been a fundamental stumbling block to understanding non-supersymmetric YM theory from first principles in a similar way, reflected practically in the absence of a direct, BCFW-like recursive computation of real-world YM integrands. 

In this letter, we propose a solution to this long-standing problem of finding ``the'' integrand for non-supersymmetric YM theory. As we will see, such an object is handed to us by the recent reformulation of YM amplitudes associated with ``surfaceology''~\cite{Arkani-Hamed:2023lbd, Arkani-Hamed:2023mvg}~\cite{Arkani-Hamed:2017mur, Arkani-Hamed:2019vag, Arkani-Hamed:2019plo, Arkani-Hamed:2019mrd}. 
The difficulty associated with single-cuts is an extremely basic, kinematical one; for instance, tadpoles are assigned zero momenta and are hence associated with a $1/0$ singularity. The resolution we offer is similarly basic and structural: we will see that the picture of the surface generalizes what we mean by kinematics, such that with the generalized objects we have ``perfect'' integrands, so that {\it e.g.} at one-loop, the loop-cut matches onto the appropriate tree-amplitude with completely generic kinematics. Furthermore, basic features of surfaceology also allow us to control the behavior at infinity, so that loop integrand can be computed as a sum over single cuts using new recursion relations unrelated to BCFW. 
\\ \\
\noindent {\bf The one-loop surface integrand.}--- The object that defines the one-loop gluon integrand, $ \mathcal{I}_n$, is the surface integral describing the scattering of $2n$ scalars from which we extract the gluon answer by taking $n$ residue~\cite{Zeros, Gluons}. In this limit, each pair of adjacent scalars fuses into a gluon and the remaining object describes what we call ``scalar-scaffolded gluons''~\cite{Gluons}:
\begin{equation}
    \mathcal{I}_n = \text{Res}_{X_{\text{Scaff}}=0} \int \prod_{P\in \mathcal{T}_S} \frac{\diff y_P}{y_P^2} \prod_{\mathcal{C}\in \mathcal{S}} u_\mathcal{C}[\{y_P\}]^{\alpha^\prime X_\mathcal{C}},
    \label{eq:SurfaceIntegrand}
\end{equation}
where $\mathcal{S}$ stands for the surface, $i.e.$ the punctured disk with $2n$ marked points on the boundary; $u_\mathcal{C}$ are the so-called $u$-variables associated with the curves $\mathcal{C}$ on the surface $\mathcal{S}$, and they appear in the surface integral raised to the power $\alpha^\prime X_\mathcal{C}$, with $X_\mathcal{C}$ the kinematic invariant associated to curve $\mathcal{C}$. The $\mathcal{T}_S$ is a triangulation for the surface $\mathcal{S}$. The $u_\mathcal{C}$'s are functions of the positive integration variables $y_P$, where each $y_P$ is associated with a propagator entering in a representative Feynman diagram of the $2n$ scalar process containing the gluon propagators, $X_{\text{Scaff}}$. This \textit{positive} parametrization $u_\mathcal{C}[\{y_P\}]$ is such that the $u$'s satisfy a set of non-trivial equations -- the so-called $u$-equations~\cite{Arkani-Hamed:2017mur, Arkani-Hamed:2019plo}~\cite{Arkani-Hamed:2023mvg, Arkani-Hamed:2023lbd}: 
\begin{equation}
    u_\mathcal{C}[\{y_P\}] + \prod_{\mathcal{C}^\prime \in \mathcal{S}} u_{\mathcal{C}^\prime}[\{y_P\}]^{\text{Int}(\mathcal{C}^\prime,\mathcal{C})}=1,
\end{equation}
where $\text{Int}(\mathcal{C}^\prime,\mathcal{C})$ is the number of times curve $\mathcal{C}^\prime$ intersects curve $\mathcal{C}$. Even though there are as many $u$ equations as $u$ variables, but since we can parametrize them in terms of $2n$ $y_P$'s~\footnote{Since the number of propagators entering on a $2n$ 1-loop diagram is $2n$, the number of integration variables $y_P$ is also $2n$.}, we conclude that the space of solutions is still $2n$ dimensional! 

Finally, in \eqref{eq:SurfaceIntegrand} we consider the product over $\mathcal{C}\in \mathcal{S}_n$ to be restricted to curves $\mathcal{C}$ that self-intersect at most \textit{once} around the puncture. Therefore we can divide this product into three parts: (I) curves starting and ending at boundary marked points/puncture with \textit{no} self-intersections, $\mathcal{C}^{(0)}$; (II) curves starting and ending at boundary marked points with one self-intersection, $\mathcal{C}^{(1)}$; and (III) the closed curve on the surface, around the puncture that we denote by $\Delta$, whose exponent we also call $\Delta$. The exponent of this closed curve is different than those of open curves, instead $\Delta$ encodes the dimensionality of spacetime, $d$, via $\Delta= 1-d$~\cite{Gluons}.

In summary, $\eqref{eq:SurfaceIntegrand}$ defines the object that at leading order at low-energies, $\alpha^\prime X_\mathcal{C} \ll 1$, gives us what we will call the \textit{surface one-loop integrand},   $\mathcal{I}^{\mathcal{S}}_n [\{X_{\mathcal{C}}\}] =\lim_{\alpha^\prime X_\mathcal{C} \ll 1} \mathcal{I}_n $, which is manifestly a rational function of the kinematic invariants $X_{\mathcal{C}}$. Even though this object starts life as a full ``stringy'' integral, in this letter we will always be studying the field-theory object $\mathcal{I}^{\mathcal{S}}_n$, describing exclusively the field-theory scalar-scaffolded gluon integrands. Nonetheless, the surface integral is crucial to derive the factorization rules that the object satisfies on both tree- and loop-cuts -- the first step towards understanding what the surface formulation is buying us, and why this object deserves to be called ``the'' integrand is to define the kinematic invariants $X_\mathcal{C}$.  
\\ \\
\noindent {\bf Surface Kinematics.}--- The surface integrands are written in terms of $X_\mathcal{C}$'s which are functions of dot products of momentum of the $2n$ scalars, $p_i \cdot p_j$. So the first thing to understand is how the gluon momenta, $q_i^\mu$, and polarizations, $\epsilon_i^\mu$, obtained after taking the scaffolding residue, are encoded in the momentum of the scalars~\cite{Gluons}.
The gluon momenta are given by $q_i^{\mu} = (p_i+p_{i+1})^\mu$, and for polarizations, since the gluons are minimally coupled to the scalars, we have $\epsilon_i^{\mu} \propto (p_i-p_{i+1})^\mu$, up to the usual gauge redundancy $\epsilon_i^{\mu} \to  \epsilon_i^{\mu} + \alpha(p_i+p_{i+1})^\mu$. One can check that in the scaffolding residue, $q_i^2=0$, we have $\epsilon_i \cdot q_i = \epsilon_i ^2 =0$.
\begin{figure}[t]
    \centering
    \includegraphics[width=\linewidth]{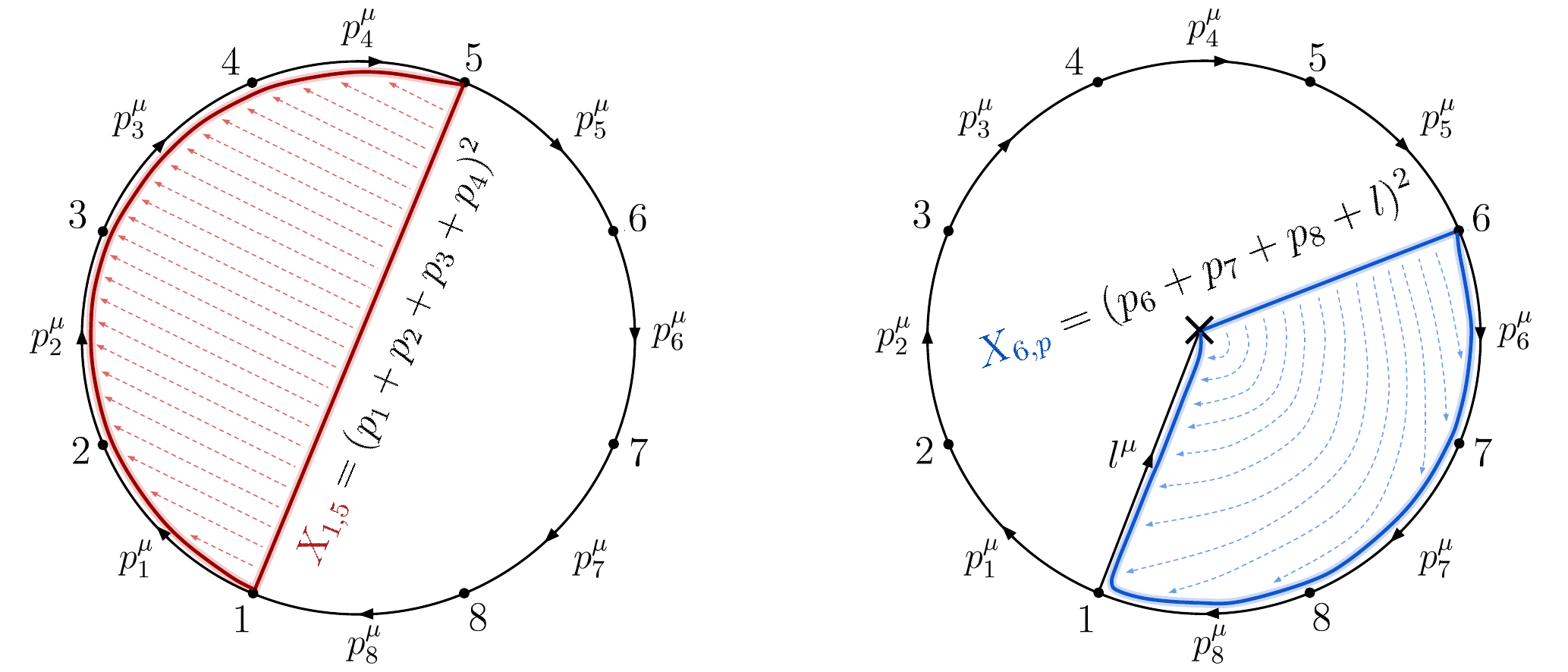}
    \caption{Determining momentum of curves via homology, at tree-level (left) and one-loop (right).}
    \label{fig:homology}
\end{figure}

Now each variable $X_\mathcal{C}$ is associated with a curve $\mathcal{C}$ on the surface $\mathcal{S}$ -- the punctured disk with $2n$ marked points on the boundary. Let us start by considering the non-self-intersecting curves, each such curve $\mathcal{C}^{(0)}$ on $\mathcal{S}$ is in one-to-one correspondence to a propagator appearing in Feynman diagrams of the $2n$-scalar process. From a curve on the surface, we can read off the momentum flowing through the corresponding propagator by \textit{homology}. At tree-level, in which case the surface is simply the disk, giving momentum $p_i^\mu$ to each boundary component, a curve from $i$ to $j$ is assigned momentum $(p_i +p_{i+1} + \cdots + p_{j-1})^\mu$, by homology. Then the $X_{\mathcal{C}}$ variable associated to this curve is simply $X_{i,j}=(p_i +p_{i+1} + \cdots + p_{j-1})^2$ (see Fig. \ref{fig:homology}, left). These are the so-called \textit{planar variables}~\cite{Arkani-Hamed:2017mur}, giving the momentum of propagators entering in planar Feynman diagrams. The scaffolding residue puts the gluons on-shell by going on the locus $(p_i + p_{i+1})^2=X_{i,i+2}=0$, with $i$ odd. So after this residue the curves that live inside the inner gluon n-gon, $X_{i,j}$ with $i$ and $j$ both odd, give us the dot products of momentum appearing in the gluon propagators, while those for which at least one of the indices is even correspond to dot products involving also the polarizations of the gluons  (see~\cite{Gluons} for a more detailed explanation of the mapping).

\begin{figure}[t]
    \centering
    \includegraphics[width=\linewidth]{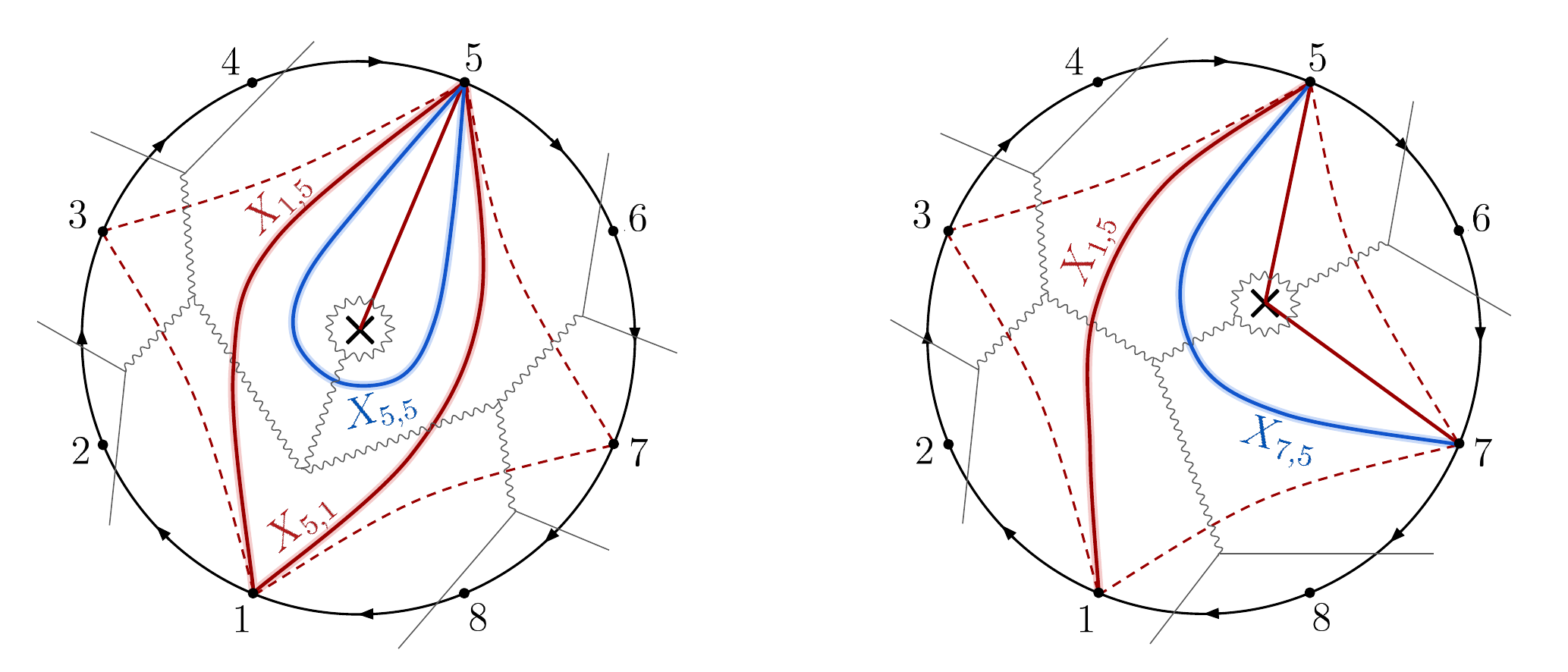}
    \caption{Tadpole (left) and external bubble (right) triangulations of the 4-point gluon one-loop surface, where each curve is assigned a different variable $X$, even when they are the same homology/have the same momentum. The scaffolding curves are dashed while the gluon internal propagators are solid lines.}
    \label{fig:TadpoleBubble}
\end{figure}

At one-loop, we can form a basis of homology by assigning momentum to each boundary of the disk, $p_i^\mu$, as well as to one of the curves ending on the puncture, $l^\mu$. Then, we can read off the momentum flowing through the dual propagators by homology. Now, we could similarly determine $X_{\mathcal{C}}$ by homology -- doing so and assuming that the puncture carries no momentum, we get that different curves on the surface are assigned the same kinematic invariant, $X$. For example, for two marked points in the boundary $i$ and $j$, we have two different curves on the surface, $X_{i,j}$, that goes through the left of the puncture, and $X_{j,i}$ going through the right, however by homology we would have $X_{i,j} = X_{j,i}$ (see fig. \ref{fig:TadpoleBubble}, left, for curves $X_{1,5}$ and $X_{5,1}$). In particular, this would imply $X_{i+1,i}=X_{i,i}=0$, as well as $X_{j+2,j}=0$ for $j$ odd, due to the scaffolding residue. Note that for $i$ odd, $X_{i,i}$ and $X_{i+2,i}$ are curves which are dual to gluon tadpoles and external bubbles propagators (see Fig. \ref{fig:TadpoleBubble}), respectively, which indeed in momentum space are assigned zero momentum. In addition, in this kinematic mapping, the self-intersecting curves are assigned the same momentum as their non-self-intersecting versions, $X_{\mathcal{C}^{(1)}} = X_{\mathcal{C}^{(0)}}$. 

Now the power of the surface integrand is precisely that it gives us the freedom to go \textit{beyond} homology, allowing us to define a generalization of kinematics -- \textit{surface kinematics} -- for which all cuts are well-defined and can be matched to gluing of lower-objects! By doing this we can bypass the standard problems of the (momentum $\leftrightarrow$ homology) situation for which single cuts are ill-defined, as explained in the introduction. 

Now let us introduce surface kinematics from which we can define an object where all the above issues are resolved. Instead of reading kinematics by homology, we assigned a different variable to each \textit{different} curve on the surface, such that $X_{i,j} \neq X_{j,i}$. However, we still assign the same variable to curves that \textit{differ by self-intersection}, except for the boundary curves $X_{i,i+1}$ which only exist for non-zero self-intersection. Therefore, for an $n$-gluon one-loop integrand, we have $(2n)^2- 2n$ $X_{i,j}$'s~\footnote{We have $X_{i,i+2}=0$ for $i$ odd from the scaffolding residue, and no $X_{a,a}$ with $a$ even can appear in the answer.}, and $2n$ $X_{i,p}$, {\it i.e.} curves ending on the puncture. It is crucial to include \textit{all} these distinct kinematic variables to be able to correctly interpret all the cuts of the integrand. In addition, as explained in~\cite{Gluons}, the surface integrand written in terms of surface kinematics, is manifestly gauge-invariant, in the most natural way gauge-invariance could be phrased in this extension of kinematics. 
\\ \\
\noindent{\bf Surface Gauge-Invariance.}--- From the surface integral, we can show that after taking the scaffolding residue the answer can be expressed as in \eqref{eq:SurfaceGaugeInv}, where the sum over $j$ ranges over indices all other than $2$, $i.e.$ $\{1,3,4,\cdots,2n,p\}$~\footnote{Of course, $X_{2,p}=X_{p,2}$ so the term $j=p$ in the sum only appears once.}. On the second line we have a correction term proportional to the boundary curve $X_{2n,1}$ multiplying derivatives of the integrand, where the second term is  evaluated with index $2\rightarrow 1$~\footnote{This means $X_{2,j} \rightarrow X_{1,j}$ and $X_{j,2} \rightarrow X_{j,1}$ for all $j$}. Just like at tree-level, gauge-invariance and multi-linearity combine such that a similar statement holds in $(X_{2,j}-X_{3,j})$, which is stated in the second equality. 

So we can see that the tree-level notion of gauge invariance and multi-linearity ports very nicely to the surface loop integrand, with the additional correction factors proportional to the boundary curves. Trivially these corrections vanish when we set the boundary curves to zero, but we choose to keep them as they make the gluing rules as simple as possible.
\\ \\
\noindent {\bf Surface Gluing Rules.}--- Directly from the surface integral 
\eqref{eq:SurfaceIntegrand}, we can derive how the cuts of the loop-surface are related to lower objects. We will here summarize the gluing rules for the two relevant cuts: the tree-tree cut and tree-loop cut, where we cut the surface via a curve $X_{a,b}$ where $a,b$ are both \textit{odd} boundary points; and the loop-cut, where we cut the surface via a curve $X_{a,p}$, where $a$ is an odd boundary point and $p$ the puncture. The reader can find a detailed derivation of these results in \cite{Gluons}.

\begin{figure*}[t]
    \centering
    \includegraphics[width=\linewidth]{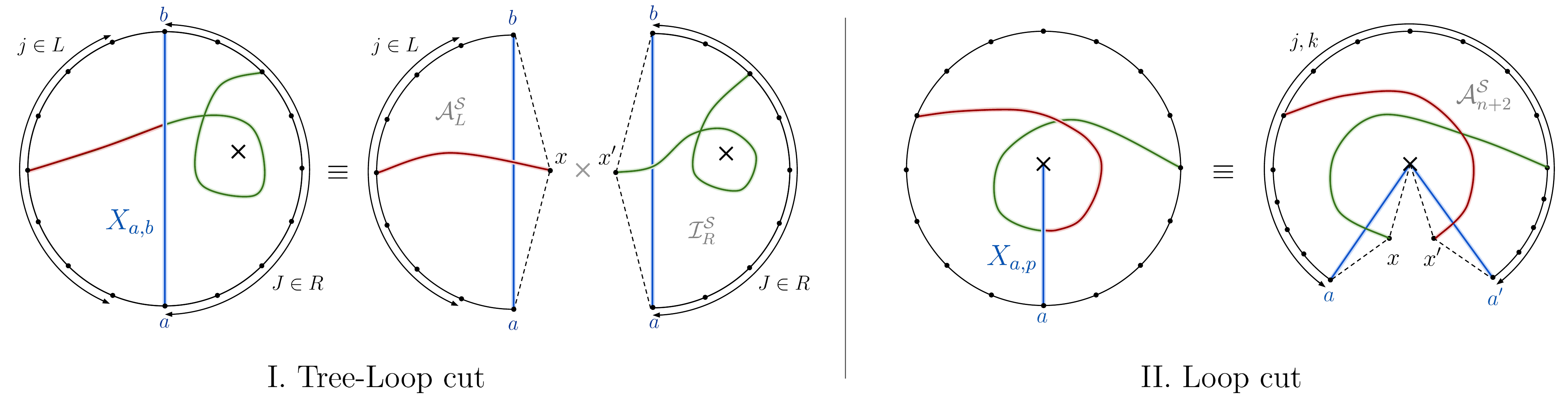}
    \caption{Different cuts of puncture disk, with curves on the original decomposing into curves on the cut surface}. 
    \label{fig:cuts}
\end{figure*}

For the tree-tree/tree-loop case, let us consider cutting the punctured disk via a curve $X_{a,b}$ (with the puncture to the right of it, in the latter case). On this cut one obtains the gluing of tree $\mathcal{A}_L^\mathcal{S}(a,a+1,\cdots,b-1,b,x)$ with another tree $\mathcal{A}_R^\mathcal{S}(b,b+1,\cdots,a-1,a,x^\prime)$ for the former, or with loop integrand $\mathcal{I}_R^\mathcal{S}(b,b+1,\cdots,a-1,a,x^\prime)$ for the latter (see Fig. \ref{fig:cuts}, left). The residue at $X_{a,b}=0$ of ${\cal A}_n^{\cal S}$ and $\mathcal{I}_n^\mathcal{S}$, is given in terms of lower-point objects as shown in eqs. \eqref{eq:tree-tree} and \eqref{eq:tree-loop}, respectively.

As for the loop-cut, we consider cutting the punctured disk along curve $X_{a,p}$, and in this limit we should obtain the gluing of a single tree $\mathcal{A}^\mathcal{S}_{n+2}(a,a+1,\cdots,a^\prime,x^\prime,p,x)$, with $a^\prime \rightarrow a$ (see Fig. \ref{fig:cuts}, right). The residue of surface integrand $\mathcal{I}_n^\mathcal{S}$, at $X_{a,p}=0$ is given in eq. \eqref{eq:loop-cut}. 
\begin{widetext}
\vspace{-5mm}
\underline{Gauge-Invariance and Linearity (in gluon 1)}:
\begin{equation}
\begin{aligned}
    \mathcal{I}_n^{\mathcal{S}} &= \sum_{j\neq 2} \left[ \left(X_{2,j} - X_{1,j}\right) \frac{\partial \mathcal{I}_n^{\mathcal{S}}}{\partial X_{2,j}} + \left(X_{j,2} - X_{j,1}\right) \frac{\partial \mathcal{I}_n^{\mathcal{S}}}{\partial X_{j,2}} \right]+ X_{2n,1} \times \left[ \frac{\partial \mathcal{I}_n^{\mathcal{S}}}{\partial X_{2n,2}} + \frac{\partial \mathcal{I}_n^{\mathcal{S}}}{\partial X_{2n,1}} \bigg \vert_{2\rightarrow 1} \right] \\
    &= \sum_{j\neq 2} \left[ \left(X_{2,j} - X_{3,j}\right) \frac{\partial \mathcal{I}_n^{\mathcal{S}}}{\partial X_{2,j}} + \left(X_{j,2} - X_{j,3}\right) \frac{\partial \mathcal{I}_n^{\mathcal{S}}}{\partial X_{j,2}} \right] + X_{3,4} \times \left[ \frac{\partial \mathcal{I}_n^{\mathcal{S}}}{\partial X_{2,4}} + \frac{\partial \mathcal{I}_n^{\mathcal{S}}}{\partial X_{3,4}} \bigg \vert_{2\rightarrow 3} \right].
\end{aligned}
\label{eq:SurfaceGaugeInv}
\end{equation}

\underline{Tree-Tree Cut}:
\begin{equation}
\mathop{\mathrm{Res}}_{X_{a,b}=0}\left(\mathcal{A}_n^{\mathcal{S}}\right) = \sum_{j\in L, J\in R\setminus \{a,b\}} \left(X_{j,J}-X_{a,J}-X_{b,j}\right) \frac{\partial \mathcal{A}_L^\mathcal{S}}{\partial X_{j,x}}\frac{\partial \mathcal{A}_R^\mathcal{S}}{\partial X_{x^\prime,J}}.
\label{eq:tree-tree}
\end{equation}

\underline{Tree-Loop Cut}:
\begin{equation}
\begin{aligned}    \mathop{\mathrm{Res}}_{X_{a,b}=0}\left(\mathcal{I}_n^{\mathcal{S}}\right) &= \sum_{j \in L, J\in R} \left[\left(X_{j,J}-X_{j,b}\hat{\delta}_{j,b-1} - X_{a,J}\right) \frac{\partial \mathcal{A}_L^\mathcal{S}}{\partial X_{j,x}}\frac{\partial \mathcal{I}_R^\mathcal{S}}{\partial X_{x^\prime,J}} + \left(X_{J,j}-X_{j,b}\hat{\delta}_{j,b-1}  - X_{J,a}\hat{\delta}_{J,a-1} \right) \frac{\partial \mathcal{A}_L^\mathcal{S}}{\partial X_{j,x}}\frac{\partial \mathcal{I}_R^\mathcal{S}}{\partial X_{J,x^\prime}}\right]\\
&\quad + \sum_{j\in L} \left[\left(X_{j,p} - X_{a,p} -X_{j,b}\hat{\delta}_{j,b-1}\right)   \frac{\partial \mathcal{A}_L^\mathcal{S}}{\partial X_{j,x}}\frac{\partial \mathcal{I}_R^\mathcal{S}}{\partial X_{x^\prime,p}}\right] + \sum_{j\in L} X_{a-1,a} \frac{\partial \mathcal{A}_L^\mathcal{S}}{\partial X_{j,x}}\frac{\partial \mathcal{I}_R^\mathcal{S}}{\partial X_{a-1,a}} \bigg\vert_{x^\prime \rightarrow a},
\end{aligned}  
\label{eq:tree-loop}
\end{equation}
with $\hat{\delta}_{i,k} = 1-\delta_{i,k}$, $L=(a+1,a+2,\cdots,b-1)$, and $R=(b,b+1,\cdots,a-1,a)$. 
\\
\par
\underline{Loop Cut}:
\begin{equation}
    \mathop{\mathrm{Res}}_{X_{a,p}=0}\left(\mathcal{I}_n^{\mathcal{S}}\right) = \sum_{j,k} \left[\left(X_{j,p} + X_{k,p} - X_{k,j} \right)\frac{\partial^2 \mathcal{A}_{n+2}^\mathcal{S}}{\partial X_{j,x}\partial X_{k,x^\prime}} \right]\bigg\vert_{a^\prime\rightarrow a} -d \frac{\partial \mathcal{A}_{n+2}^{\mathcal{S}}}{\partial X_{x^\prime,x}}; \quad j,k \in (a,a+1,\cdots,a-1, a^\prime).
    \label{eq:loop-cut}
\end{equation}
\end{widetext}
\noindent {\bf Vanishing at Infinity and Recursion.}--- Now that we have an object where all the cuts are well defined, to recursively construct it, we just need to find a direction for which it vanishes at infinity. This turns out to be extremely simple, and is once more directly given to us by the general properties of the surface integral, as will be explained (in concert with a number of other new properties of scalar, pion and gluon amplitudes at infinity) in \cite{cexp}. Here we simply summarize the results needed for the recursive determination of tree amplitudes and one-loop integrands. 

We begin at tree-level. As explained previously, for YM theory we always pick the triangulations defining the surface to contain the scaffolding curves, $X_{\text{scaff}}$, and allow any curves to appear inside the gluonic n-gon. Now, given any triangulation, every curve $X$ on the surface is associated with a ``$g$-vector'', ${\bf g}_X$. It is then natural to shift the kinematics as $X \to X + z ({\bf g}_X \cdot {\bf t})$, where ${\bf t}$ is any fixed direction we like, and we imagine varying $z$. The claim is that under this shift, if the direction ${\bf t}$ is generic enough so that all the poles are shifted, i.e. so that ${\bf g} \cdot t \neq 0$ for all $X$, then the tree amplitude vanishes as $z \to \infty$, which is not entirely trivial, since the tree amplitude has units of $X^2$, and so individual terms in the amplitude blow up as $z^2$ as $z \to \infty$, but has a simple power-counting explanation 
\footnote{This shift preserves all $c_{i,j}$'s for at least one of $i,j$ being odd, which  correspond to Lorentz products involving gluon polarizations~\cite{Gluons}, thus by power-counting, the YM amplitude goes like $1/z^{n{-}3{-}\lfloor \frac n 2\rfloor}$ as $z\to \infty$.}.
This vanishing is reminiscent of the soft behavior at infinity under BCFW shifts, though ours are not BCFW shifts. By the usual Cauchy theorem argument, we can then express the amplitude recursively as a sum over the residues on all the shifted poles. As usual, each term in the recursion relation will have spurious poles, which can be seen to non-trivially cancel in the full sum, giving a check on the consistency of the picture. 

The same story holds at one-loop. Here there is an especially natural $g-$ vector shift, where we simply uniformly shift all the loop-variables, as $X_{i,p} \to X_{i,p} + z$, where $z$ is the same for all $i$. The integrand again vanishes at infinity. This is again a non-trivial statement: the one-loop integrand is dimensionless, and term-by-term, we have pieces with equal numbers of $X_{i,p}$ in the numerator and denominator. Because this shift only touches loop momenta, the resulting recursion relation expresses the integrand as a sum over single-loop-cuts, each of which via the expressions given above are determined by tree amplitudes. Again each term in the recursion relation has spurious poles in the form of $1/(X_{i,p} - X_{j,p})$, that non-trivially cancel in the full sum. As another check, while the integrand is constructed purely from the knowledge of the single-loop-cuts, all the other ``tree $\times$ one-loop'' cuts can be shown to exactly match using the gluing rules defined above. 

Having recursively obtained the integrand for generalized surface kinematics, we can of course go back to the usual momentum-kinematics given by homology. This will produce $1/0$ divergences from tadpoles and external bubbles, which are guaranteed to multiply scaleless integrals that integrate to zero. So we can either remove all such terms manually, or we can do the loop integrals in general $d$ dimension first, and then specialize the surface kinematics to the usual ones. We have checked that through four points, the integrated amplitudes obtained in this way match the known expressions in the literature near four dimensions, for all helicities. 
\\ \\
\noindent {\bf Outlook.}---
Having a canonical notion of ``the'' loop-integrand for non-supersymmetric theories directly relevant to the real world opens the door to exploring a new world of ideas. We have emphasized that these ``perfect'' objects manifest fundamental properties such as gauge-invariance and cuts. But their connection to surfaceology suggests they have many further properties -- hidden zeros~\cite{Zeros,Bartsch:2024amu,Li:2024qfp,Rodina:2024yfc}, splits patterns and multi-soft limit~\cite{Splits} (see also~\cite{Cao:2024gln,Cao:2024qpp,Zhang:2024iun})
, and connections to the NLSM and Tr($\phi^3$) theory~\cite{Arkani-Hamed:2024nhp}, further properties at infinity to be discussed in \cite{cexp}, and beyond. We close by discussing some obvious directions for immediate development. 

In this letter we have defined the one-loop integrand and determined it recursively, but nothing in our ideas is restricted to one-loop, and we expect the methods to be extended to determining non-supersymmetric planar YM integrands at all loop orders (see also~\cite{Bern:2024vqs} for recent work on constructing loop integrands at all loop orders). As familiar in surfaceology the only new feature at two loops and above is modding out by the mapping class group. This poses no new difficulties for the YM case. As an illustration, in the appendix we show how single cuts as well as vanishing at infinity under g-vector shifts can be used to determine the two-loop tadpole integrand, using the most obvious method for modding out by the mapping class group. We expect this to extend straightforwardly to all multiplicity. Furthermore, for Tr($\phi^p$) theory there is a simple way of determining all-loop planar surface integrands solving certain (essentially combinatorial) differential equations~\cite{Arkani-Hamed:2024pzc}, that should extend naturally to our YM setting. 

The generalization of kinematics and cutting/gluing rules came directly from the--naturally ``stringy''--surface integral. Nonetheless, the pure-field theoretic integrands we have found are autonomous objects, which exhibit striking regularities not found in their $\alpha^\prime$ extensions. Might there be a sort of ``amplituhedron'' that directly computes these integrands in a new way, not connected to the surface integrals? 

We have stressed our methods are inherently dimension-agnostic since $d$ appears simply as a parameter in the gluing rules defining the cuts. Relatedly, the expressions for tree and loop integrands are ``longer'' than those we are accustomed to seeing in (fixed-helicity) sectors of gluon amplitudes in four dimensions. It would clearly be fascinating to find a more intrinsically four-dimensional version of surface kinematics.  
\\ \\
\noindent {\bf Acknowledgements.---}
The work of N.A.H. is supported by the DOE (Grant No. DE-SC0009988), the Simons Collaboration on Celestial Holography, the ERC UNIVERSE+ synergy grant, and the Carl B. Feinberg cross-disciplinary program in innovation at the IAS. The work of Q.C. is supported by the National Natural Science Foundation of China under Grant No. 123B2075. The work of C.F. is supported by FCT/Portugal
(Grant No. 2023.01221.BD). The work of S.H. has been supported by the National Natural Science Foundation of China under Grant No. 12225510, 11935013, 12047503, 12247103, and by the New Cornerstone Science Foundation through the XPLORER PRIZE.

\onecolumngrid
\appendix
\section{Recursion Relations and Examples}
In this appendix, we present the recursion relations for tree amplitudes and one-loop integrands based on surface kinematics, as well as explicit examples computed from them. In both cases, we reconstruct the rational function based on its residues which are in turn given by the surface gluing rules described in the main text, which thus leads to a recursion for tree amplitude in terms of lower-point ones, and that for one-loop integrand in terms of tree amplitudes. 

\subsection{Tree recursion and examples}
At tree level, we seek for a kinematics shift with good behavior at infinity, which allows us to reconstruct the amplitude recursively by residue theorem. 
It turns out that the four-point amplitude is special in that there is no $g$-vector shift with vanishing large-$z$ behavior, but such ``contact term" at infinity can be computed from the surface integral. For $n\geq5$, the shift associated with any ``scaffolding triangulation" has good behavior for $z\to \infty$.  

To be concrete, let us choose the inner-gon triangulation to be the ray-like triangulation (all the $X_{1,a}$ with $a$ odd). In the $2n$-gon kinematics, we keep the $X_{\text{scaff}}=X_{a,a+2}$ unshifted by setting ${\bf g}_{X_{\text{scaff}}} \cdot {\bf t}=0$, and shift all the $X$ variables of the inner-gon along the ray-like direction, which is induced by $X_{1,a}\to X_{1,a}+z(t_{1,a})$ where we denote ${\bf g}_{X_{1,a}} \cdot {\bf t}\equiv t_{1,a}$. One can use the path rule to calculate the $g$-vector for each curve~\cite{Arkani-Hamed:2023lbd}. Explicitly, all $X_{i,j}$ are shifted as follows,
\begin{equation}
    X_{i,j}\to X_{i,j}+z(- t_{1,2\lceil \frac{i}{2}\rceil +1}+ t_{1,2\lceil \frac{j}{2}\rceil -1})\,,
\end{equation}
where recall that $t_{1,3}$, $t_{1,n-1}$ are zero for keeping $X_{\text{scaff}}$ unshifted. For example, for $n=4$ in addition to $X_{1,5}\to X_{1,5}+ z t_{1,5}$, $X_{1,6}, X_{2,5}, X_{2,6}$ are also shifted by $z t_{1,5}$; we have $X_{3,7}\to X_{3,7}-z t_{1,5}$ and also $X_{3,8}, X_{4,7}, X_{4,8}$ shifted by $-z t_{1,5}$ (all other $X$ variables unshifted). 

In this way, all the poles of the tree amplitude are shifted and we have checked up to $n=8$ that the amplitude vanishes at $z\to  \infty$ for $n\geq 5$ (one should be able to derive such large-$z$ behaviour from the surface integral directly but we leave this to the future work). With the knowledge of the large-$z$ behaviour, we can use the Cauchy theorem to determine the tree amplitude: 
\begin{equation}
	\mathcal{A}_n^{\mathcal{S}}=-\sum_{z=z^{\star}} \mathop{\mathrm{Res}}_{X_{a,b}(z^\star)=0}\frac{\mathcal{A}_n^{\cal S}(z^{\star})}{z^\star}=\sum_{(a,b)} \frac{\mathop{\mathrm{Res}}_{X_{a,b}=0}\left(\mathcal{A}_n^{\mathcal{S}}(X(z^{\star}))\right)} {X_{a,b}}({\bf g}_{X_{a,b}} \cdot {\bf t})\,,
\end{equation}
where all the poles are given by $X_{a,b}(z)=X_{a,b}+z({\bf g}_{X_{a,b}} \cdot {\bf t})=0$, thus $z^{\star}=-X_{a,b}/({\bf g}_{X_{a,b}} \cdot {\bf t})$, with $a,b$ odd. The residue at $X_{a,b}(z^{\star})=0$ is given by that of $\mathcal{A}_n^{\mathcal{S}}$ at $X_{a,b}=0$ with all other $X \to X+z^{\star}{\bf g}_{X} \cdot {\bf t} =X-X_{a,b}\frac{{\bf g}_{X} \cdot {\bf t}}{ {\bf g}_{X_{a,b}} \cdot {\bf t}}$, and very nicely it is given by the tree-tree cut~\eqref{eq:tree-tree} (gluing of two lower-point tree amplitudes), with shifted variables $X\to X(z^{\star})$. Therefore, this gives a recursion relation which constructs the tree amplitude for $n\geq 5$ in terms of lower-point amplitudes. In fact for $n=4$, since we know the non-vanishing terms at $z\to \infty$ from the surface integral we can still use this formula by explicitly adding the residue at infinity. For example, for the inner-gon triangulation with $X_{1,5}$, we get the recursion relation as,
\begin{equation}
	\mathcal{A}_4^{\mathcal{S}} = \frac{\mathop{\mathrm{Res}}_{X_{1,5}=0}\left(\mathcal{A}_4^{\mathcal{S}}(X(z=-\frac{X_{1,5}}{t_{1,5}}))\right)} {X_{1,5}/t_{1,5}}-\frac{\mathop{\mathrm{Res}}_{X_{3,7}=0}\left(\mathcal{A}_4^{\mathcal{S}}(X(z=\frac{X_{3,7}}{t_{1,5}}))\right)} {X_{3,7}/t_{1,5}}+\mathcal{A}_4^{\mathcal{S}}(z\to \infty)\,,
\end{equation}
where the pole $X_{3,7}$ is shifted to $X_{3,7}+z({\bf g}_{X_{3,7}} \cdot {\bf t})=X_{3,7}-z t_{1,5}$, so the second term is evaluated as the residue at $X_{3,7}(z)=0$. The third term is from the residue of the pole at infinity, which turns out to be simply  $c_{1,5}c_{3,7}$ (recall $c_{i,j}:=X_{i,j}+X_{i{+}1, j{+}1}-X_{i,j{+}1}-X_{i{+}1, j}$)! The upshot is that we obtain exactly the same result as computed in~\cite{Gluons}:
\begin{equation}\label{eq:four-point amp}
{\cal A}_4^{\cal S}=\frac{N_{1,5}}{X_{1,5}}+ \frac{N_{3,7}}{X_{3,7}}+ c_{1,5} c_{3,7}  
\end{equation}
with the ``numerator" $N_{1,5}$ expressed in terms of $c$'s as
\begin{equation}
    \begin{aligned}
      N_{1,5}= &-c_{3,5678}\left(c_{1,67}c_{5,7}+c_{1,5} c_{12345,7}+c_{1,7} c_{5,8}\right)-c_{1,4} \left(c_{12345,7} c_{3,5}+c_{3,67} c_{5,7}+c_{3,7} c_{5,8}\right)\\
        &-c_{1,3} \left(c_{123,5}c_{1234,7}+c_{123,567}c_{5,7}+c_{123,7}c_{5,8}\right).
    \end{aligned}
\end{equation}
and $N_{3,7}=N_{1,5}|_{i\to i{+}2}$, where we have introduced the shorthand notation $c_{A,B} = \sum_{k \in A, l \in B} c_{k,l}$ so that e.g. $c_{12,34} = c_{1,3} + c_{1,4} + c_{2,3} + c_{2,4}$. Expanding the expression in terms of $X$'s we have
\begin{equation}
	\begin{aligned}
	{\cal A}_4^{\cal S}=&\frac{X_{1,4} X_{1,6} X_{2,7}-X_{1,4} X_{1,6} X_{2,8}+X_{1,4} X_{1,6} X_{3,8}-X_{1,6} X_{2,4} X_{3,8}}{X_{1,5}}-\frac{X_{1,6} X_{2,5} X_{4,8}}{2 X_{1,5}}\\
	&+\frac{X_{1,4} X_{2,7} X_{3,6}-X_{1,4} X_{2,8} X_{3,6}+X_{2,7} X_{3,8} X_{4,6}+X_{2,5} X_{3,8} X_{4,7}}{X_{3,7}}-\frac{X_{2,7} X_{3,6} X_{4,8}}{2 X_{3,7}}\\
	&+(X_{1,5}-X_{1,4}-X_{1,6}) (X_{2,7}-X_{2,8})+X_{1,6} (X_{3,7}-X_{3,8})+(i\to i+2,i+4,i+6)\\
	&+ X_{2,6} X_{4,8}-X_{1,5} X_{3,7}-X_{2,8} X_{4,6}-X_{2,4} X_{6,8}\\
    &-\frac{X_{1,5} \left(X_{2,7}-X_{2,8}+X_{3,8}\right) \left(X_{3,6}-X_{4,6}+X_{4,7}\right)}{X_{3,7}}-\frac{X_{3,7} \left(X_{1,4}-X_{2,4}+X_{2,5}\right) \left(X_{1,6}+X_{5,8}-X_{6,8}\right)}{X_{1,5}},
	\end{aligned}
\end{equation}
where $(i\to i+2,i+4,i+6)$ denotes terms obtained by cyclic shifts of $\{1,2, \cdots,8\}$ by $2,4,6$ for all preceding terms. 

For $n\geq 5$ the recursion requires no contribution from $z\to \infty$, and we have computed up to $7$-point tree amplitudes in terms of surface kinematics, which are in the ancillary file. 

\subsection{One-loop recursion and examples} 
At one-loop, we shift all the loop-variables as $X_{i,p} \to X_{i,p} + z$, and we expect $\frac{{\cal I}^{\cal S}_n(z)}{z}$ vanishes as $z\to \infty$ so that we can reconstruct the one loop integrand $\mathcal{I}_n^{\mathcal{S}}$ from its residues at the finite position by the Cauchy theorem:
\begin{equation}
	\mathcal{I}_n^{\mathcal{S}}=-\sum_{z=z^{\star}} \mathop{\mathrm{Res}}_{X_{i,p}(z^\star)=0}\frac{\mathcal{I}_n^{\cal{S}}(z^{\star})}{z^\star}=\sum_{i} \frac{\mathop{\mathrm{Res}}_{X_{i,p}=0}\left(\mathcal{I}_n^{\mathcal{S}}(X_{j,p}\to X_{j,p}-X_{i,p})\right)} {X_{i,p}}\,,
\end{equation}
where the poles of are at $X_{i,p}(z)=X_{i,p}+z=0$ for odd $i$, and the residue at $X_{i,p}(z^{\star})=0$ is given by that of $\mathcal{I}_n^{\mathcal{S}}$ at $X_{i,p}=0$ with all the $X_{j,p}\to X_{j,p}+z^{\star}=X_{j,p}-X_{i,p}$ for $j\neq i$. As we have discussed, all these residues are given by the loop cut~\eqref{eq:loop-cut}, which is the gluing of tree amplitudes with $n{+}2$ legs (with shifted kinematics $X_{j,p}\to X_{j,p}-X_{i,p}$).  

For example, the two-point integrand ${\cal I}_2^{\cal S}$ can be obtained in this way, with the two residues given in terms of four-point tree computed above:
\begin{equation}
		\mathcal{I}_2^{\mathcal{S}}= \frac{\mathop{\mathrm{Res}}_{X_{1,p}=0}\left(\mathcal{I}_2^{\mathcal{S}}(X_{j,p}\to X_{j,p}-X_{1,p})\right)} {X_{1,p}}+\frac{\mathop{\mathrm{Res}}_{X_{3,p}=0}\left(\mathcal{I}_2^{\mathcal{S}}(X_{j,p}\to X_{j,p}-X_{3,p})\right)} {X_{3,p}}\,,
\end{equation}
and the spurious pole $X_{3,p}-X_{1,p}$ is cancelled in the final result. Explicitly we find (here we denote $X_{i,p}\equiv Y_i$ for simplicity):
\begin{equation}
\begin{aligned}
    {\cal I}_2^{{\cal S}}=   &-\frac{X_{1,2} X_{1,4}+X_{2,1} X_{4,1}+(d{-}2)(X_{2,1}X_{1,4}-Y_2 X_{1,4}-Y_4 X_{2,1}+Y_3\left(X_{1,4}+X_{2,1}-X_{2,4}\right))}{Y_1 X_{1,1}}\\
    &+\frac{X_{1,2}-X_{2,1}+X_{4,1}-X_{1,4}+2 X_{2,4}-X_{4,2}+(d{-}2)(Y_2-Y_3+Y_4-X_{2,4})}{Y_1}+\frac{X_{2,4}}{X_{1,1}}
    \\&+\frac{Y_2 \left(X_{1,4}-X_{4,1}+X_{4,3}-X_{3,4}\right)+X_{1,1} \left(X_{3,2}-X_{4,2}+X_{4,3}\right)}{Y_1 Y_3}+(i\to i+2)\\
    &-\frac{X_{1,4} X_{3,2}+X_{2,1} X_{4,3}+X_{1,1} X_{3,3}+(d{-}2)Y_2 Y_4}{Y_1 Y_3}-(d{-}2)\,,
\end{aligned}
\end{equation}
where $(i\to i+2)$ denotes the cyclic permutation $\{1,2,3,4\}\to \{3,4,1,2\}$ for all preceding terms, and those terms on the last line are manifestly invariant under $(i\to i{+}2)$.

We have also computed the integrand ${\cal I}^{\cal S}_n$ explicitly from recursion up to $n=5$, which we record in the ancillary file. 

\subsection{Consistency checks} The results we have obtained pass numerous consistency checks: as we have mentioned, all spurious poles are canceled in both ${\cal A}_n^{\cal S}$ and ${\cal I}_n^{\cal S}$, and one can check good large-$z$ behavior of our results explicitly for ${\cal I}_n^{\cal S}$ and for ${\cal A}_n^{\cal S}$ with $n\geq 5$. Moreover, we have checked surface gauge-invariance and linearity for tree amplitudes and one-loop integrands. For example, the $n=2$ integrand can be written as gauge invariant and linear in gluon $1$ by \eqref{eq:SurfaceGaugeInv}:
\begin{equation}
{\cal I}_2^{{\cal S}}=\sum_{j=1,3,4,p}(X_{2,j}-X_{1,j}){\cal Q}_{X_{2,j}} + \sum_{j=1,3}(X_{j,2}-X_{j,1}){\cal Q}_{X_{j,2}} 
+ X_{4,2} {\cal Q}_{X_{4,2}}+X_{4,1} {\cal Q}'_{X_{4,1}}\,,
\end{equation}
where ${\cal Q}_X:=\frac{\partial {\cal I}_{2}^{\cal S}}{\partial X}$ except for the last boundary term ${\cal Q}'_{X_{4,1}}:=\frac{\partial {\cal I}_{2}^{\cal S}}{\partial X_{4,1}}|_{2\to 1}$, and they read
\begin{equation}
\begin{aligned}
&{\cal Q}_{X_{2,1}}=\frac{X_{3,3}-X_{4,3}+Y_4}{Y_1 Y_3}-\frac{X_{4,1}}{Y_1 X_{1,1}}-\frac{1}{Y_1}+(d-2)\frac{X_{1,4}+Y_3-Y_4}{Y_1 X_{1,1}}\,,\\
&{\cal Q}_{X_{1,2}}=-\frac{X_{1,4}}{Y_1 X_{1,1}}-\frac{Y_4}{Y_1 Y_3}+\frac{1}{Y_1}\,,\\
&{\cal Q}_{X_{2,3}}=-\frac{X_{4,3}}{Y_3 X_{3,3}}-\frac{Y_4}{Y_1 Y_3}+\frac{1}{Y_3}\,,\\
&{\cal Q}_{X_{3,2}}=\frac{X_{1,1}-X_{1,4}+Y_4}{Y_1 Y_3}-\frac{X_{3,4}}{Y_3 X_{3,3}}-\frac{1}{Y_3}+(d-2)\frac{X_{4,3}+Y_1-Y_4}{Y_3 X_{3,3}}\,,\\
&{\cal Q}_{X_{2,4}}=-\frac{X_{3,3}+Y_1-2 Y_3}{Y_1 Y_3}+(d-2)\frac{-X_{1,1}+Y_1-Y_3}{Y_1 X_{1,1}}\,,\\
&{\cal Q}_{X_{4,2}}=\frac{-X_{1,1}+2 Y_1-Y_3}{Y_1 Y_3}+(d-2)\frac{-X_{3,3}-Y_1+Y_3}{Y_3 X_{3,3}}\,,\\
&{\cal Q}_{Y_2}=\frac{X_{1,4}-X_{4,1}+X_{4,3}-X_{3,4}}{Y_1 Y_3}+(d-2)\left(-\frac{X_{1,4}}{Y_1 X_{1,1}}-\frac{X_{4,3}}{Y_3 X_{3,3}}-\frac{Y_4}{Y_1 Y_3}+\frac{1}{Y_1}+\frac{1}{Y_3}\right)\,,\\
&{\cal Q}'_{X_{4,1}}=-\frac{1}{Y_3}\,.
\end{aligned}
\end{equation}

Furthermore, even though we have constructed surface loop integrands from loop cuts \eqref{eq:loop-cut}, they also satisfy \eqref{eq:tree-loop} for tree-loop cuts as they should, and we have explicitly checked this for our results. Last but not least, we can also integrate our loop integrands and compare them with known results for one-loop amplitudes ({\it c.f.}~\cite{Bern:1991aq, Bern:1992em, Bern:1994zx}. After reducing {\it e.g.} ${\cal I}^{\cal S}_4$ to a basis of integrals in $d=4- 2 \epsilon$, we obtain the correct integrated gluon amplitudes for all helicity configurations. 

\section{Tadpole Cut}
The tadpole cut is a special type of tree-loop cut for which $X_{a,b} \equiv X_{a,a}$, and so $R=\{a\}$. In this case, the tree/loop factors are, respectively, $\mathcal{A}_L^\mathcal{S}(a,a+1,\cdots,a-1,a,x)$ and $\mathcal{I}_R^\mathcal{S}(a,x^\prime)$, and the expression for the cut reduces to:
\begin{equation}
\begin{aligned}
\mathop{\mathrm{Res}}_{X_{a,a}=0}\left(\mathcal{I}_n^{\mathcal{S}}\right) &= \sum_{j\in L} \left(X_{a,j} -X_{j,a}\hat{\delta}_{j,a-1}\right) \frac{\partial \mathcal{A}_L^\mathcal{S}}{\partial X_{j,x}}\frac{\partial \mathcal{I}_R^\mathcal{S}}{\partial X_{a,x^\prime}} +  \left(X_{j,p} -X_{j,a}\hat{\delta}_{j,a-1} -X_{a,p}\right) \frac{\partial \mathcal{A}_L^\mathcal{S}}{\partial X_{j,x}}\frac{\partial \mathcal{I}_R^\mathcal{S}}{\partial X_{x^\prime,p}} \\
& \quad + X_{a-1,a}\frac{\partial \mathcal{A}_L^\mathcal{S}}{\partial X_{a-1,x}}\frac{\partial \mathcal{I}_R^\mathcal{S}}{\partial X_{x^\prime,a}},  
\end{aligned}
\end{equation}
with $\hat{\delta}_{i,j} = 1-\delta_{i,j}$ and $L = \{a+1,\cdots,a-1\}$.
\section{Two-Loop Tadpole}
Here we show how the recursion discussed in the text for one-loop generalizes to higher loops, by studying the one-point two-loop integrand, $\mathcal{I}_1^{\mathcal{S}_2}$, with $\mathcal{S}_2$ being the disk with two punctures. For simplicity, we will focus only on the part proportional to $d^2$, $i.e.$ the dimension of spacetime squared. This is the simplest part of the integrand but it already contains all the non-trivialities that start at two-loops, namely dealing with the infinite winding curves coming from the action of the mapping class group (MCG), which just corresponds to a single winding at two-loops. We will begin by using cuts and good behavior at infinity under $g$-vector shift to compute the ``infinite integrand", which sums over all winding copies of the integrand we want. We will then see how surface gauge-invariance allows us to canonically represent the infinite integrand we obtain from recursion in a form that makes it trivial to mod out by winding and get the planar integrand we seek. 

The procedure we will follow has three steps: 
\begin{enumerate}
    \item We will start by finding a direction for which the infinite integrand vanishes at infinity. This direction will be such that we only shift poles that give us planar contributions that we can compute using the 1-loop results we derived in the rest of the paper.
    \item Having found a direction for which the integrand vanishes at infinity, we can apply Cauchy's theorem to write it as a sum over the shifted poles. This sum is then infinite precisely because we are dealing with the infinite integrand where we keep all the windings. The spurious poles appearing in this sum all cancel telescopically, and the final object has no spurious poles. 
    \item Finally we are left with an object which corresponds to infinite copies of the integrand, once we identify curves in the same MCG orbit. To extract a single copy, we show that the surface gauge-invariance statement for the integrand allows us to write it in a manifestly MCG invariant form, from which we can extract the integrand (post-modding out by the MCG). 
\end{enumerate} 

\subsection{The two-loop ``g-vector'' shift}

\begin{figure}[t]
    \centering
    \includegraphics[width=\linewidth]{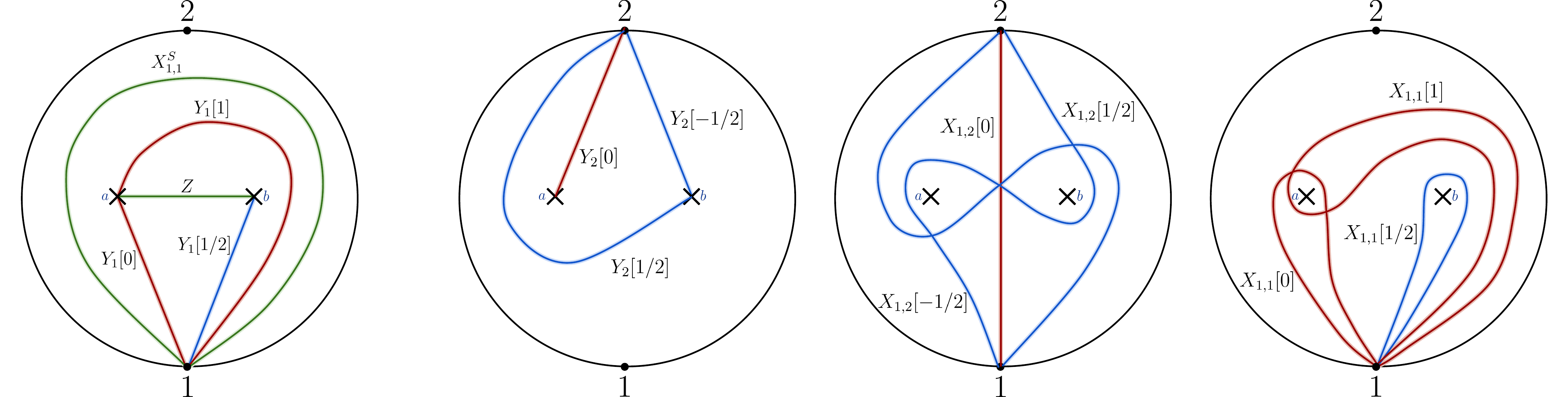}
    \caption{(Left) Base triangulation for the one-point two-loop integrand. In red we have the curves with integer winding, in blue those with half-integer winding and in green those that don't have winding. (Right) Examples of different curves and their respective windings. }
    \label{fig:2loops}
\end{figure}

Let's start with point 1., and derive the ``g-vector'' shift for the two-loop recursion. We are computing the one-point two-loop gluon integrand, so we start with a two-point scaffolded problem, and we consider the triangulation of $\mathcal{S}_2$ -- the two punctured disk with two marked points on the boundary -- containing the scaffolding curve $X^S_{1,1}$ that goes around both curves, as well as  $Z$ -- the curve connecting one puncture to the other -- and curves $Y_{1}[0]$, $Y_{1}[+1/2]$, $Y_{1}[+1]$ (see figure \ref{fig:2loops}), where the number inside the brackets stands for the winding: $+1$,$+1/2$ corresponding to a $360^{\circ}$, $180^{\circ}$ rotation of the punctures counter-clockwise, respectively. So all curves $X[n]$ with $n\in \mathbb{Z}$ belong to the same MCG class, and similarly those with $X[n+1/2]$ with $n\in \mathbb{Z}$. For simplicity, let's call the puncture on the left $a$ and the one on the right $b$, so that both $Y_1[n]$ start in $1$ and end in $a$ while $Y_1[n+1/2]$ starts in $1$ and ends in $b$. 

Since we have to take the scaffolding residue on $X^S_{1,1} \to 0$, it is important not to shift this variable, so we set  $t^S_{1,1}=0$. The cut of any variable that is shifted will appear in our recursion. If we cut on the curve $Z$ connecting the two punctures, or on $X_{a,a},X_{b,b}$ which are tadpoles anchored to the punctures $a,b$, we open the surface to factors containing an annulus. There is nothing wrong with this in principle, but for our purposes here we only want to use the single-trace one-loop integrands we have computed in the body of the letter, and therefore we choose that these curves are also not shifted. This is achieved by choosing $\textbf{g}_{Z} \cdot \textbf{t} =0$ and $\textbf{g}_{Y_1[i]} \cdot \textbf{t} =1$ for $i=0,1/2,1$.

This uniquely the shift of the relevant remaining curves to be:
\begin{equation}
\begin{aligned}
   & X_{1,1}[0] \to X_{1,1}[0], \\
   & X_{1,1}[+n] \to X_{1,1}[+n]+2z, \\
   & X_{1,1}[-n] \to X_{1,1}[-n]-2z, \\
   & X_{1,1}[n-1/2] \to X_{1,1}[n-1/2]+2z, \\
   & X_{1,1}[-n+1/2] \to X_{1,1}[-n+1/2]-2z, \\ 
   \\ 
   & Y_{1}[0] \to Y_{1}[0] + z, \\
   & Y_{1}[+n] \to Y_{1}[+n]+z, \\
   & Y_{1}[-n] \to Y_{1}[-n]-z, \\
   & Y_{1}[n-1/2] \to Y_{1}[n-1/2]+z, \\
   & Y_{1}[-n+1/2] \to Y_{1}[-n+1/2]-z, \\
\end{aligned} \quad \quad  
\begin{aligned}
   & X_{1,2}[0] \to X_{1,2}[0], \\
   & X_{1,2}[+n] \to X_{1,2}[+n]+2z, \\
   & X_{1,2}[-n] \to X_{1,2}[-n]-2z, \\
   & X_{1,2}[n-1/2] \to X_{1,2}[n-1/2]+2z ,\\
   & X_{1,2}[-n+1/2] \to X_{1,2}[-n+1/2]-2z ,\\ 
   \\
    & Y_{2}[0] \to Y_{2}[0] + z, \\
   & Y_{2}[+n] \to Y_{2}[+n]+z, \\
   & Y_{2}[-n] \to Y_{2}[-n]-z, \\
   & Y_{2}[n-1/2] \to Y_{2}[n-1/2]+z, \\
   & Y_{2}[-n+1/2] \to Y_{2}[-n+1/2]-z, \\
\end{aligned} \quad \quad \quad \text{with }n\in\mathbb{N}^+,
\label{eq:gvector-2loops}
\end{equation}
where $X_{1,2}[0]$ stands for the curve going from marked point one to two through the middle of the punctures without any winding, and $X_{1,2}[n]$ its respective winding versions. $X_{1,1}[0]$ is then the tadpole going around puncture $a$ while $X_{1,1}[1/2]$ goes around puncture $b$, and similarly $Y_{2}[0]$ starts in $2$ and ends in puncture $a$ while $Y_{2}[-1/2]$ ends in puncture $b$. Under this shift and when we send $z\to \infty$ the infinite surface integrand, $\mathcal{I}_1^{\mathcal{S}_2}$ vanishes. 

\subsection{Summing over cuts}

From the shift derived in the previous section, we see that the poles that are shifted correspond to curves $X_{1,1}$ and $Y_1$, since we know the gluon integrand obtained after scaffolding residue does not have poles in $X_{1,2}$ and $Y_2$. So to compute the two-loop integrand, we need to derive these two cuts: the loop-cut, $Y_1[n]=0$, and the loop-loop cut, $X_{1,1}[n]=0$. As we mentioned earlier, for the sake of this appendix, we will only apply the recursion to compute the part of the integrand proportional to $d^2$, $\mathcal{I}_1^{\mathcal{S}_2,d^2}$, and in particular, this means that in the loop-cut, \eqref{eq:loop-cut}, we only want to keep the piece proportional to $d^2$, 
 $\mathcal{I}_1^{\mathcal{S}_2,d^2}$, so that for zero winding curve, $Y_{1}[0]$ at two-loops we get:
\begin{equation}    \text{Cut}_{Y_1[0]}=\mathop{\mathrm{Res}}_{Y_{1}[0]=0}\left(\mathcal{I}_1^{\mathcal{S}_2,d^2}\right) = - \left( \frac{\partial \mathcal{I}_3^{\mathcal{S}_1}}{\partial X_{x,x^\prime}} + \frac{\partial \mathcal{I}_3^{\mathcal{S}_1}}{\partial X_{x^\prime,x}} \right) \bigg\vert_{d},
\label{eq:loop-cut-d2}
\end{equation}
where $\mathcal{I}_3^{\mathcal{S}_1} \equiv \mathcal{I}_3^{\mathcal{S}_1}(1,2,1^\prime,x^\prime,a,x)$, is the three-point one loop integrand we obtain when we cut the two-loop surface open via curve $Y_1[0]$, and $a$ stands for puncture $a$. Now the curves on $\mathcal{I}_3$ entering the cut \eqref{eq:loop-cut-d2} can be identified as curves on $\mathcal{S}^2$ according to the following map 
\begin{equation}
\begin{aligned}
    &Y_1 \to Y_1[-1/2], \\
    &Y_{a} \to Z, \\
    &Y_{1^\prime} \to Y_1[1/2], \\
    &Y_2 \to Y_2[-1/2], \\
\end{aligned} \quad \quad \quad 
\begin{aligned}
    &X_{1,1} \to X_{1,1}[-1/2], \\
    &X_{1,a} \to Y_1[-1], \\
    &X_{2,a} \to Y_2[-1],\\
    &X_{a,2} \to Y_2[0], \\
\end{aligned}\quad \quad \quad 
\begin{aligned}
    &X_{a,1^\prime} \to Y_{1}[+1], \\
    &X_{1^\prime,1} \to X_{1,1}[0], \\
    &X_{a,a} \to X_{a,a},\\
    &X_{1^\prime,1^\prime} \to X_{1,1}[+1/2], \\ 
\end{aligned} \quad \quad \quad 
\begin{aligned}
    &X_{2,1} \to X_{1,2}[-1/2], \\
    &X_{1^\prime,2} \to X_{1,2}[0]. \\
\end{aligned}
\label{eq:kinMap-LoopCut}
\end{equation}

Now if instead we were interested in the cut at $Y_1[n]$ for some $n \neq 0$, (integer or half-integer), the expression \eqref{eq:loop-cut-d2} still holds, except now in the kinematic mapping \eqref{eq:kinMap-LoopCut} we must add $n$ to all the windings appearing.

In addition, we need to derive the tadpole cut, and from it extract the $d^2$ part. Once more, let's do this for the zero winding curve, $X_{1,1}[0]$. In the residue, we expect the result to be given by the gluing of the two lower-objects: $\mathcal{I}_1^{\mathcal{S}_1}(1,x)$ and $\mathcal{I}_2^{\mathcal{S}_1}(1^\star,2,1^\prime,x^\prime)$ Following the same strategy as explained in \cite{Gluons}, we can derive the residue of the surface integral, $\mathop{\mathrm{Res}}_{X_{1,1}[0]=0}\left(\mathcal{I}_1^{\mathcal{S}_2,d^2}\right)$, to be \footnote{To make it simpler we are omitting the contributions coming from the self-intersecting boundary curves, $X_{1,2}^{(1)}$ and $X_{2,1}^{(1)}$, as these don't give any contribution to the $d^2$ part of the integrand.}:
\begin{equation}
\begin{aligned}
 \text{Cut}_{X_{1,1}[0]} =  &\left[\left(Z-Y_1[0]-Y_1[+1/2]\right) \frac{\partial \mathcal{I}_1^{\mathcal{S}_1} }{Y_x} \frac{\partial \mathcal{I}_2^{\mathcal{S}_1} }{Y_{x^\prime}}+ \left(Y_2[0]-Y_1[0]-X_{1,2}[0]\right)\frac{\partial \mathcal{I}_1^{\mathcal{S}_1} }{Y_x} \frac{\partial \mathcal{I}_2^{\mathcal{S}_1} }{X_{x^\prime,2}} \right. \\
    &\left.\quad + \left(Y_2[-1]-Y_1[0]\right)\frac{\partial \mathcal{I}_1^{\mathcal{S}_1} }{Y_x} \frac{\partial \mathcal{I}_2^{\mathcal{S}_1} }{X_{2,x^\prime}}+  \left(Y_1[1]-Y_1[0]\right) \frac{\partial \mathcal{I}_1^{\mathcal{S}_1} }{Y_x} \frac{\partial \mathcal{I}_2^{\mathcal{S}_1} }{X_{x^\prime,1^\prime}}+  \left(Y_1[-1]-Y_1[0]\right) \frac{\partial \mathcal{I}_1^{\mathcal{S}_1} }{Y_x} \frac{\partial \mathcal{I}_2^{\mathcal{S}_1} }{X_{1^\star,x^\prime}} \right. \\
    &\left. \quad + X_{1,2}[-1/2] \frac{\partial \mathcal{I}_1^{\mathcal{S}_1} }{X_{1,x}} \frac{\partial \mathcal{I}_2^{\mathcal{S}_1} }{X_{2,x^\prime}} - X_{1,2}[0] \frac{\partial \mathcal{I}_1^{\mathcal{S}_1} }{X_{1,x}} \frac{\partial \mathcal{I}_2^{\mathcal{S}_1} }{X_{x^\prime,2}} - Y_{1}[0] \frac{\partial \mathcal{I}_1^{\mathcal{S}_1} }{Y_{x}} \left(\frac{\partial \mathcal{I}_2^{\mathcal{S}_1} }{X_{x^\prime,1^\star}}+\frac{\partial \mathcal{I}_2^{\mathcal{S}_1} }{X_{1^\prime,x^\prime}} \right)  \right. \\
    &\left. \quad -Y_1[1/2]\left(\frac{\partial \mathcal{I}_1^{\mathcal{S}_1} }{X_{x,1}}+\frac{\partial \mathcal{I}_1^{\mathcal{S}_1} }{X_{1,x}} \right)\frac{\partial \mathcal{I}_2^{\mathcal{S}_1} }{Y_{x^\prime}} -X_{1,1}[1/2]\left(\frac{\partial \mathcal{I}_1^{\mathcal{S}_1} }{Y_{x}}+\frac{\partial \mathcal{I}_1^{\mathcal{S}_1} }{X_{x,1}}+\frac{\partial \mathcal{I}_1^{\mathcal{S}_1} }{X_{1,x}} \right) \left(\frac{\partial \mathcal{I}_2^{\mathcal{S}_1} }{X_{x^\prime,1^\prime}}+\frac{\partial \mathcal{I}_2^{\mathcal{S}_1} }{X_{1^\prime,x^\prime}} \right)  \right] \bigg \vert_{d^2},
\end{aligned}
\label{eq:tadpole-cut-d2}
\end{equation}
and similarly the curves of $\mathcal{I}_1^{\mathcal{S}_1}$ and $\mathcal{I}_2^{\mathcal{S}_1}$ can be mapped into curves in $\mathcal{I}_2^{\mathcal{S}_2}$ according to:
\begin{equation}
\underline{\mathcal{I}_1^{\mathcal{S}_1}:}\quad  \begin{aligned}
&Y_1 \to Y_1[0], \\
\end{aligned} \quad \quad \underline{\mathcal{I}_2^{\mathcal{S}_1}:}\quad  \begin{aligned}
&Y_{1^\star} \to Y_1[-1/2], \\
&Y_{1^\prime} \to Y_1[+1/2], \\
&Y_{2} \to Y_2[-1/2] ,
\end{aligned} \quad \quad \begin{aligned}
&X_{1^\prime,1^\prime} \to X_{1,1}[+1/2], \\
&X_{1^\star,1^\star} \to X_{1,1}[-1/2], \\
\end{aligned} \quad \quad \begin{aligned}
&X_{1^\prime,2} \to X_{1,2}[0], \\
&X_{2,1^\star} \to X_{1,2}[-1/2]. \\
\end{aligned} 
\label{eq:KinMap-tadpole}
\end{equation}

Just like before, if instead we want to get the residue at $X_{1,1}[n]$ for any $n$ integer or half-integer, all we need to do is use \eqref{eq:tadpole-cut-d2} with the mapping \eqref{eq:KinMap-tadpole} where we add $n$ to all the windings entering the expressions. Note that when we shift by a half-integer, the (non-winding) curve $X_{a,a}$ is mapped to $X_{b,b}$ and vice-versa. 

Now that we have all the cuts, we can write the integrand as a sum over the shifted poles according to \eqref{eq:gvector-2loops} which leads to:
\begin{equation}
\begin{aligned}
    \mathcal{I}_1^{\mathcal{S}_2,d^2} =&\frac{\text{ Cut}_{Y_1[0]} \vert_{z= -Y_{1}[0]}}{Y_1[0]} + \sum_{n \in \frac{1}{2}\mathbb{Z}^+}\frac{\text{Cut}_{Y_1[n]} \vert_{z= -Y_{1}[n]}}{Y_{1}[n]} + \frac{\text{Cut}_{X_{1,1}[n]} \vert_{z= -X_{1,1}[n]/2}}{X_{1,1}[n]} \\
    & +\sum_{n \in \frac{1}{2}\mathbb{Z}^-}\frac{\text{Cut}_{Y_1[n]} \vert_{z= Y_{1}[n]}}{Y_1[n]} + \frac{\text{Cut}_{X_{1,1}[n]} \vert_{z= X_{1,1}[n]/2}}{X_{1,1}[n]},
    \end{aligned}
\end{equation}
where each cut depends on $z$ via the g-vector shift, $X\to X+z\textbf{g}_X\cdot t$, defined earlier.

Note that the terms in this sum over winding cuts all have spurious poles involving adjacent winding. For instance the cut on $Y1[2]$ have poles for the and on $Y1[1] + z$ and $Y1[3]+z$, which evaluated at $z = - Y1[2]$ become spurious poles in $(Y1[1] - Y1[2]), (Y1[3] - Y1[2])$. These spurious poles cancel telescopically in the sum, as as strong consistency check on the computation. 

\subsection{Modding out by winding to extract the planar integrand}

The infinite sum over residues gives us the ``infinite integrand", which sums over an infinite number of copies of the actual planar integrand we want, differing only by windings. We wish instead to identify all the variables in a given winding sector, but doing this on the infinite integrand would gives us the planar integrand we like multiplied by an infinite factor. What we should do instead is mod out by the action of the mapping class group. In this case, where the mapping class group is trivially winding, this is in principle especially easy to do. It must be possible to put the infinite integrand in the form $I = \sum_{w = -\infty}^\infty F_w$, where $F$ only depends on $w$ through the labels of the curves, but otherwise has exactly the same form for all winding. This makes the invariance under shifts $w \to (w+1)$ manifest. To mod out by winding, we just take the term from one winding sector in this sum, say $F_0$, and then identify all the variables. But while our recursion is given as a sum over all windings, it is {\it not} directly of this form, since the form of the result for positive, zero, and negative windings are different, reflecting the inevitable dependence of the $g$-vector shifts on our choice of base triangulation. 

But very nicely, surface gauge invariance guarantees that that there is a canonical way we can write the integrand, which is precisely of the form we need. According to surface gauge-invariance, we can write the $d^2$ part of the integrand as follows:
\begin{equation}
\begin{aligned}
     \mathcal{I}_1^{\mathcal{S}_2,d^2} = &\sum_{n \in \mathbb{Z}} (Y_2[n]-Y_1[n])\frac{\partial \mathcal{I}_1^{\mathcal{S}_2,d^2}}{\partial Y_2[n]} + (X_{1,2}[n]-X_{1,1}[n])\frac{\partial \mathcal{I}_1^{\mathcal{S}_2,d^2}}{\partial X_{1,2}[n]}\\
     &+(Y_2[n+1/2]-Y_1[n+1/2])\frac{\partial \mathcal{I}_1^{\mathcal{S}_2,d^2}}{\partial Y_2[n+1/2]} + (X_{1,2}[n+1/2]-X_{1,1}[n+1/2])\frac{\partial \mathcal{I}_1^{\mathcal{S}_2,d^2}}{\partial X_{1,2}[n+1/2]},
\end{aligned}
\end{equation}
which makes manifest the fact that the full infinite integrand is invariant under the MCG, $X[n]\to X[n+1]$.
Using this fact, we can pick a single term in this sum, say the one with zero winding, and identify all curves in the same MCG class to obtain the final answer: 
\begin{equation}
\begin{aligned}
     \mathcal{I}_1^{\mathcal{S}_2,d^2} = &\left[(Y_2[0]-Y_1[0])\frac{\partial \mathcal{I}_1^{\mathcal{S}_2,d^2}}{\partial Y_2[0]} + (X_{1,2}[0]-X_{1,1}[0])\frac{\partial \mathcal{I}_1^{\mathcal{S}_2,d^2}}{\partial X_{1,2}[0]} \right.\\
     &\left.+(Y_2[1/2]-Y_1[1/2])\frac{\partial \mathcal{I}_1^{\mathcal{S}_2,d^2}}{\partial Y_2[1/2]} + (X_{1,2}[1/2]-X_{1,1}[1/2])\frac{\partial \mathcal{I}_1^{\mathcal{S}_2,d^2}}{\partial X_{1,2}[1/2]} \right]_{X[n]\to X[0],X[n+1/2]\to X[1/2]}.
\end{aligned}
\end{equation}

By doing this, we obtain the following expression for $ \mathcal{I}_1^{\mathcal{S}_2,d^2}$:
\begin{equation}
\begin{aligned}   \mathcal{I}_1^{\mathcal{S}_2,d^2}=&\frac{Z X_{1,2}[0]}{Y_1[0] Y_1[1/2] X_{1,1}[0]
   X_{1,1}[1/2]}+\frac{Z X_{1,2}[1/2]}{Y_1[0]
   Y_1[1/2] X_{1,1}[0]
   X_{1,1}[1/2]}+\frac{Y_2[1/2] Z}{Y_1[0]
   Y_1[1/2]{}^2 X_{1,1}[0]}-\frac{2 Z}{Y_1[0]
   Y_1[1/2] X_{1,1}[0]}\\
   &+\frac{Y_2[0] Z}{Y_1[0]{}^2
   Y_1[1/2] X_{1,1}[1/2]}-\frac{2 Z}{Y_1[0]
   Y_1[1/2] X_{1,1}[1/2]}-\frac{X_{1,2}[0]}{Y_1[0]
   X_{1,1}[0]
   X_{1,1}[1/2]}-\frac{X_{1,2}[0]}{Y_1[1/2]
   X_{1,1}[0]
   X_{1,1}[1/2]}\\
   &-\frac{X_{1,2}[1/2]}{Y_1[0]
   X_{1,1}[0]
   X_{1,1}[1/2]}-\frac{X_{1,2}[1/2]}{Y_1[1/2] X_{1,1}[0] X_{1,1}[1/2]}+\frac{2 Y_2[0]}{Y_1[0]
   Y_1[1/2] X_{1,1}[0]}-\frac{2 Y_2[1/2]}{Y_1[0]
   Y_1[1/2]
   X_{1,1}[0]}\\
   &-\frac{Y_2[1/2]}{Y_1[1/2]{}^2
   X_{1,1}[0]}+\frac{3}{Y_1[0] X_{1,1}[0]}-\frac{2 Y_2[0]}{Y_1[0]
   Y_1[1/2] X_{1,1}[1/2]}-\frac{Y_2[0]}{Y_1[0]{}^2
   X_{1,1}[1/2]}+\frac{2 Y_2[1/2]}{Y_1[0]
   Y_1[1/2]
   X_{1,1}[1/2]}\\
   &+\frac{3}{Y_1[1/2]
   X_{1,1}[1/2]}+\frac{Y_1[1/2] Y_2[0]}{X_{a,a}
   Y_1[0]{}^2 Z}-\frac{Y_2[1/2]}{X_{a,a} Y_1[0]
   Z}-\frac{Y_2[0]}{X_{a,a} Y_1[0]{}^2}+\frac{1}{X_{a,a}
   Y_1[0]}-\frac{Y_2[0]}{X_{b,b} Y_1[1/2] Z}\\
   &+\frac{Y_1[0]
   Y_2[1/2]}{X_{b,b} Y_1[1/2]{}^2
   Z}-\frac{Y_2[1/2]}{X_{b,b}
   Y_1[1/2]{}^2}+\frac{1}{X_{b,b}
   Y_1[1/2]}+\frac{Y_2[0]}{Y_1[0]{}^2
   Z}+\frac{Y_2[1/2]}{Y_1[1/2]{}^2
   Z}-\frac{1}{Y_1[0] Z}-\frac{1}{Y_1[1/2]
   Z}\\
   &+\frac{Y_2[0]}{Y_1[0]{}^2
   Y_1[1/2]}+\frac{Y_2[1/2]}{Y_1[0]
   Y_1[1/2]{}^2}-\frac{2}{Y_1[0] Y_1[1/2]}.
\end{aligned}
\end{equation}

Note that for the infinite integrand directly coming from the recursion, spurious pole cancellation is telescopic, so that the full infinite sum is needed to cancel all poles. By contrast when we use surface gauge invariance, a given $Y_2[n],Y_2[n+1/2],X_{1,2}[n],X_{1,2}[n+1/2]$ term only occurs in the winding sectors adjacent to $n$! So the spurious poles all cancel manifestly. Using gauge invariance in this form then means that we never encounter any infinite sums in the computation of the final integrand -- we only need to keep sectors in the neighborhood of e.g. zero winding. It is striking that in this way, surface gauge invariance makes it easier to control the infinite integrand for Yang-Mills, than the analagous computation just for Tr($\phi^3$) theory. These features should carry over to all multiplicity at two-loops, and offer a new way of thinking about modding out by the MCG for Yang-Mills integrand at all loop orders.

\bibliographystyle{apsrev4-1.bst}
\bibliography{Refs.bib}

\end{document}